# RADIO-PHOTOLUMINESCENCE OF HIGHLY IRRADIATED LiF:Mg,Ti AND LiF:Mg,Cu,P DETECTORS

A.Mrozik[1], P.Bilski[1], B.Marczewska[1], B.Obryk[1], K.Hodyr[2], W.Gieszczyk[1]

1-Institute of Nuclear Physics PAN, Radzikowskiego 152, 31-342 Krakow, Poland
2-Institute of Applied Radiation Chemistry, Lodz University of Technology, Wroblewskiego 15, 93-120 Lodz, Poland

**HIGHLIGHTS**

The RPL readouts of LiF detectors were performed for ultra-high doses.

Linearity range of LiF RPL signal was investigated.

Thermal stability of RPL signal and its connection to CCs are showed.

Application of combined RPL/TL is proposed to improve high-dose dosimetry readouts.

**ABSTRACT**

The radio-photoluminescent (RPL) characteristics of LiF:Mg,Ti (MTS) and LiF:Mg,Cu,P (MCP) thermoluminescent detectors, routinely used in radiation protection dosimetry, were investigated after irradiation with ultra-high electron doses ranging up to 1 MGy. The photoluminescence of both types of LiF detectors was stimulated by a blue light (460 nm) and measured within a spectral window around 530 nm. The RPL dose response was found to be linear up to 50 kGy and sublinear in the range of 50 kGy to 1 MGy for MCP detectors and linear up to 3 kGy and next sublinear in the range from 5 kGy to 1 MGy for MTS detectors. For both type of LiF detectors RPL signal is saturated for doses higher than 100 kGy. The observed differences between MCP and MTS may suggest, that the RPL effect in LiF is not entirely governed by intrinsic defects ($F_2$ and $F_3^+$ centers), but impurities may also have a significant influence. Due to the non-destructive character of the RPL measurement, it is suggested to apply combined RPL/TL readouts, what should improve accuracy of high-dose dosimetry.

# INTRODUCTION

The photoluminescence of irradiated lithium fluoride (LiF) can be observed after stimulation with blue light. The emission spectrum of LiF, both pure and doped but containing color centers (CCs) produced by ionizing radiation, consists of two broad bands at 530 nm connected with $F_3^+$ centers and 640 nm related to $F_2$ centers (Baldacchini, 2002, Oster *et al*., 2011). Radio-photoluminescence (RPL) may be used as a non-destructive method for dose estimation, because detectors can be readout an infinite number of times with the detector signal remaining constant. Application of this phenomena was proposed by Murphy (Murphy *et al*., 2003 a,b), who described high-dose film dosimeters based on pure LiF emitting 535 nm light by excitation with 440 nm light. Detectors of commercial name Sunna were composed of a LiF powder mixed with a polymer.

LiF is also widely used as thermoluminescent (TL) detectors. Few years ago, Bilski and Obryk (Bilski *et al*., 2008; Obryk *et al*., 2009; Obryk *et al*., 2010; Obryk *et al*., 2011a) described high temperature emission of LiF:Mg,Cu,P detectors, which were heated up to 600 °C after exposure to high dose in the range from 0.5 kGy to 1 MGy. The main dosimetric peak (220 °C at the heating rate 2 °C/s ) starts to disappear with increasing doses and the new peaks start to grow up in higher temperature, which are not visible after irradiation with doses lower than 1 kGy. This new peak was called 'peak B' and the UHTR (*ultra-high temperature ratio*) method was developed and applied for dosimetry of high-doses (Obryk *et al*., 2011b).

Determination of ultra-high doses is necessary around high energy accelerators, thermo-nuclear fusion technology facilities, to dosimetry in food and medical sterilization and also in accident dosimetry.

The aim of this work was to investigate the RPL properties of the well-known LiF-based thermoluminescent detectors: LiF:Mg,Cu,P (MCP) and LiF:Mg,Ti (MTS) after high-dose irradiation in perspective of applying the dual readout, RPL and TL on a single LiF detector for achieving better accuracy of high-dose dosimetry. The work was dealing also with the study of the thermal stability of RPL signal and its connection to CCs.

**MATERIALS AND METHODS**

Two types of LiF detectors: Mg,Ti and Mg,Cu,P manufactured at the Institute of Nuclear Physics in Krakow were investigated in this work. The detectors had the shape of pellets with a diameter of 4.5 mm and a thickness of 0.9 mm. All the samples before irradiation were annealed: MTS for 1 h at 400 °C and 2 h at 100°C and MCP for 10 minutes at 260 °C and for 10 minutes at 240 °C.

Irradiation of detectors was carried out with electron beam from Linear Electron Accelerator ELU-6e installed at the Institute of Applied Radiation Chemistry, Lodz University of Technology. Detectors were irradiated with doses ranging from 0.5 kGy to 1 MGy.

The radio-photoluminescence of MCP and MTS detectors was measured in the nondestructive way using the HELIOS - 2 reader (Mandowski *et al*., 2010; Marczewska *et al*., 2012). This is a self-constructed, portable reader, which allows the measurements of RPL signal of LiF detectors, where the detectors are excited using blue diode (460 nm). PL reader has Hamamatsu H8259 "photon counting" gating photomultiplier and interference filters produced by Knight Optical Company, for excitation 460FIB12 and for emission 532FIB25, 550FIR25 and 550FIW25.

HELIOS - 2 has two available modes of stimulation: continuous-wave mode and pulsed mode. All measurements were carried out in pulsed mode. RPL readout cycle was performed with the following parameters: the stimulation time was 5000 ms, the readout time was 10 ms divided into 100 readout frames and the cycles were repeated 100 times. The scheme of the measurement cycle in the boxcar pulse mode is shown in Piaskowska *et al*., 2013.

The pre-heat procedure was performed using Harshaw 3500 TL reader at the temperature range of 50 – 270 °C and 50 – 370 °C with the heating rate of 2 °C/s for MCP and MTS detectors, respectively.

**RESULTS AND DISCUSSION**

The signals for both type detectors after irradiation with doses of 1 kGy, 10 kGy and 100 kGy are shown in Fig. 1.

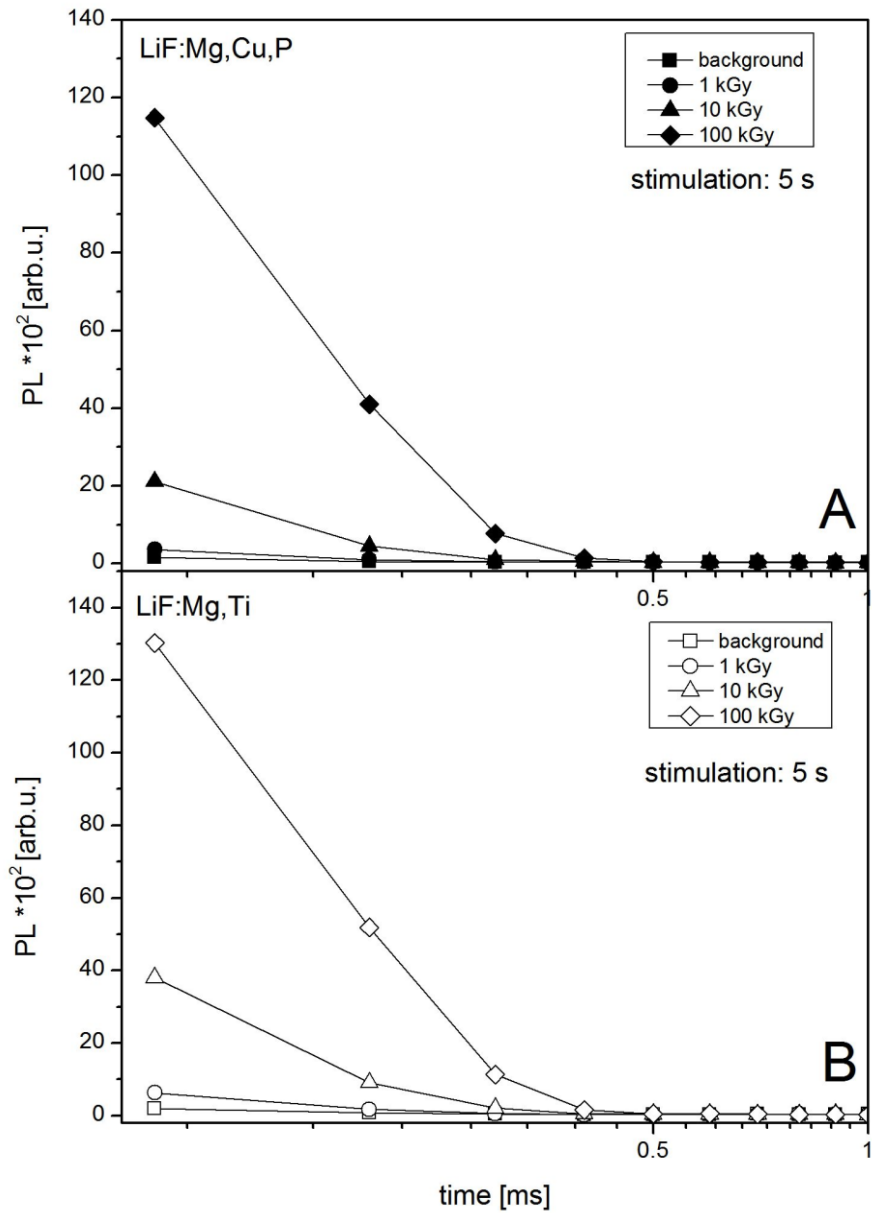

*Fig. 1. Time-resolved RPL signal after 5 s blue stimulation of LiF:Mg,Cu,P (A) and LiF:Mg,Ti (B) detectors irradiated with doses of 1 kGy, 10 kGy, 100 kGy.*

The information about the dose is collected within the first 0.5 ms and then the signal remains on the same level independently on dose. The signals for both type of detectors have different maximum level for different doses.

Fig. 2 presents the RPL dose response for MCP and MTS detectors.

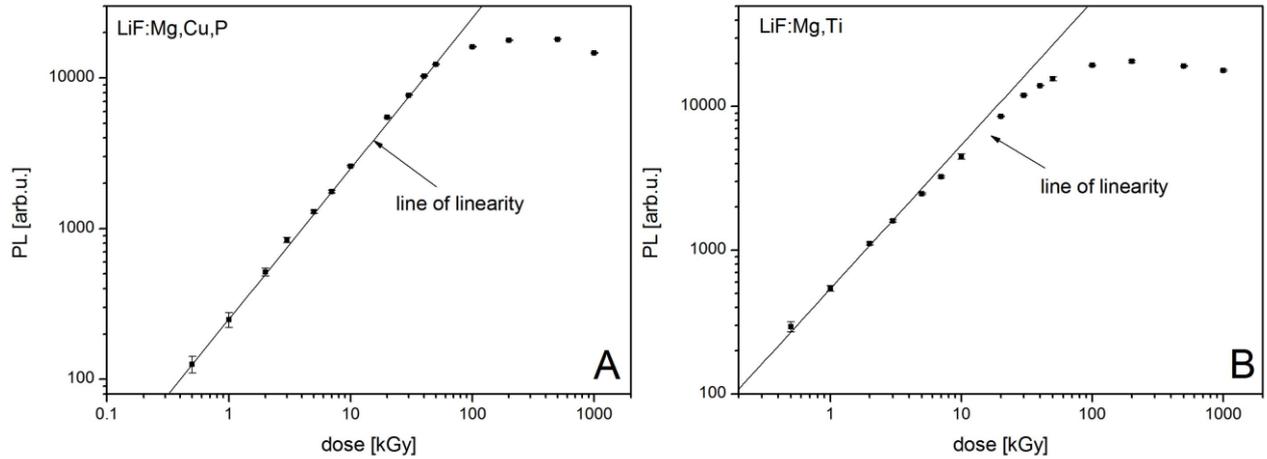

*Fig. 2. RPL signal as a function of the dose in the range of 0.5 – 1 MGy for LiF:Mg,Cu,P (A) and LiF:Mg,Ti (B) detectors.*

The background value measured for not irradiated detectors was subtracted from the RPL signals. For LiF:Mg,Cu,P detectors the dose response was found to be linear up to 50 kGy. For doses in the range from 50 kGy to 1 MGy MCP detectors are sublinear. One can see that RPL signal is saturated for doses higher than 100 kGy and starts to decrease for doses higher than 0.5 MGy. The dose characteristic is completely different for MTS detectors. The data show that the dose response is linear up to ca 3 kGy and next start to be sublinear in the range of 5 kGy – 1MGy. However, for MTS detectors, dose response above 100 kGy seems to be similar in nature to MCP detectors, namely, RPL signal becomes saturated. The dose response is often expressed in terms of linearity index:

$$f(D) = \frac{I(D)/D}{I(D_0)/D_0} \quad , \text{ where}$$

I – the intensity of RPL signal

D – irradiated dose

$D_0$ – dose within linear range of dose response

The measured data were normalized to the dose of 1 kGy. All results are presented in Fig. 3. The results illustrate stronger sublinearity of MTS than of MCP in the range above 50 kGy. Dashed line indicates linear trends.

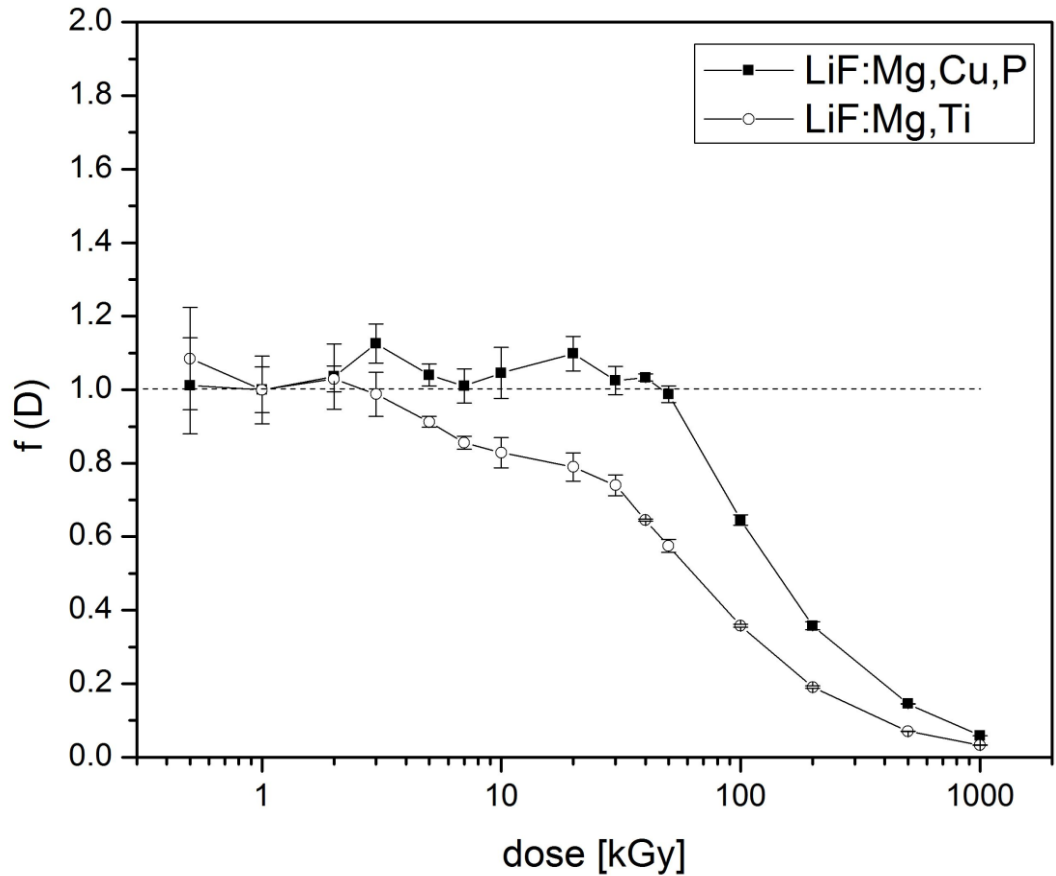

Fig. 3. Linearity index of LiF:Mg,Cu,P and LiF:Mg,Ti detectors.

Thermal stability of the RPL signal is presented in Fig. 4 for LiF:Mg,Cu,P detectors. Measurements were made for both types of detectors.

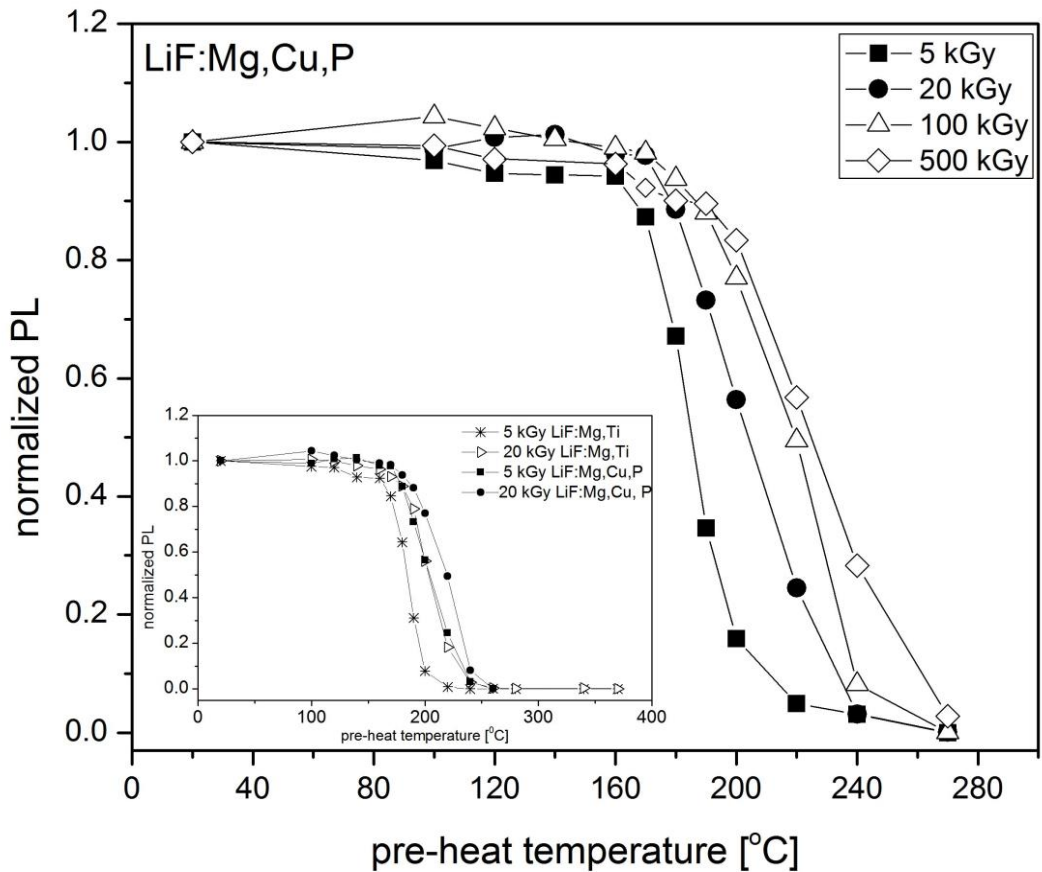

*Fig. 4. RPL signal as a function of pre-heat temperature for LiF:Mg,Cu,P detectors irradiated with different doses. Comparison of RPL signals for LiF:Mg,Cu,P and LiF:Mg,Ti detectors irradiated with the same dose (inset graph).*

In the first step, the detectors were pre-heated up to fixed temperature in the thermoluminescent reader and then the measurement was conducted in the RPL reader. The process of decreasing RPL signal begins above 160 °C independently of absorbed dose. RPL intensity is gradually decreased to the background level in the range of temperature from 60 °C to 270 °C for high-dose. A half-drop of the RPL signal (50% loss of signal) takes place at different pre-heat temperatures dependently on absorbed dose. With the increasing dose, the pre-heat temperature has to be higher in order to achieve half-drop level of signal.

Inset graph (Fig. 4) shows the comparison between thermal stability of RPL signal for MCP and MTS detectors irradiated with different doses. The measured data show difference between rate of emptying traps for both type detectors which are irradiated with the same dose. The MCP detectors need higher temperature to reach half-drop level of signal, which were irradiated with the same dose as MTS detectors. It can be associated with different type of dopants in LiF.

The RPL signal is completely removed at 270 °C independently on absorbed dose in the whole studied dose range from 0.5 kGy to 1 MGy. This finding differs from results obtained for low doses, where the RPL signal was found to reach the background level already at 220 °C (Piaskowska *et al.*, 2013). MCP and MTS detectors, which were exposed to high-dose radiation, are colored and the color of detectors is changed more and more with the increasing dose (Obryk *et al.*, 2011a). Our experiment with detectors pre-heated to different temperature showed that the detectors recover their natural color after heating to 270 °C, similarly as their RPL signal is completely lost. This confirms correlation between RPL and occurrence of color centers in LiF (Baldacchini *et al.*, 2002).

As was mentioned, the photoluminescent signal of LiF produces two broad emission bands at ~ 530 nm (connected with $F_3^+$ centers) and ~ 640 nm (corresponding with $F_2$ centers). Due to the configuration of filters used in the Helios-2 reader, only $F_3^+$ emission was measured.

Baldacchini *et al.* (2008) attributed many of the thermoluminescent glow peaks, that appear at different temperatures, to the annealing of corresponding CCs. According to their findings, the TL process starts with decay of $F_3^+$ centers, what is shown as a GP (glow peak) at 164 °C, also $F_2$ centers are ascribed to a GP at 263 °C. However, this picture may be too simple, as it does not take into account a possible role of dopants. Our results reveal strong influence of doping on RPL characteristics (the difference between MCP and MTS detectors).

Fig. 5 shows comparison of PL signal as a function of pre-heat temperature and TL glow curves for LiF:Mg,Cu,P detectors irradiated with dose of 5 kGy and 0.5 MGy. The main TL GP for 5 kGy is around 220 °C and peak B (visible above 50 kGy) for 0.5 MGy above 400 °C. As mentioned earlier, RPL signal started to disappear around 160 °C and is not visible above 270 °C. It can mean that different mechanisms and different centers are responsible for the RPL and TL phenomena for MCP.

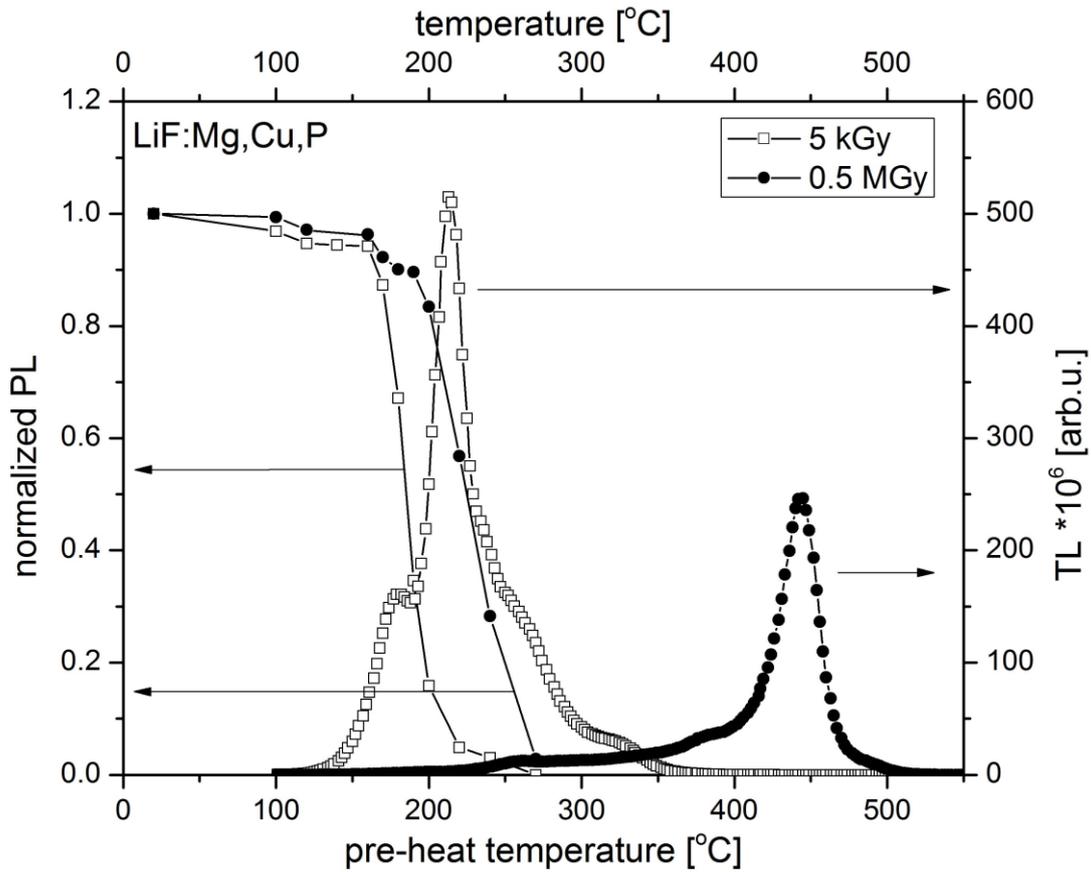

*Fig. 5. Comparison of RPL signal as a function of pre-heat temperature and TL glow curve for LiF:Mg,Cu,P detectors irradiated with doses of 5 kGy and 0.5 MGy.*

**CONCLUSIONS**

Studies of photoluminescence of highly irradiated LiF:Mg,Cu,P and LiF:Mg,Ti detectors were performed after their irradiation with the electron beam. LiF:Mg,Cu,P and LiF:Mg,Ti detectors were irradiated with doses up to 1 MGy. The RPL signal is saturated for doses exceeding 100 kGy and starts to decrease above 0.5 MGy for both type of detectors. In the case of MCP detectors, the RPL dose response was found to be linear in wide range of doses from 0.5 to 50 kGy but is sublinear above this range. MTS detectors show that the dose response is linear up to 3 kGy and in the range from 5 kGy to 1 MGy is sublinear. The RPL signal starts to decrease above 160 °C independently of absorbed dose. The half-drop signal is dependent on the absorbed dose and also LiF dopants. The observed differences between MCP and MTS may suggest that the RPL effect in LiF is not entirely governed by intrinsic defects, but impurities may also have a significant influence. The color of detector, which was irradiated with ultra-high dose, is lost after thermal treatment to 270 °C, what is associated with the total signal loss. The RPL signal disappears independently on TL peak position, which indicates that two different mechanisms are responsible for RPL and TL phenomena.

Due to the non-destructive character of the RPL measurement it may be successfully applied in combination with a TL measurement. The RPL readout prior to the TL readout would provide an independent evaluation of the dose, thus increasing accuracy of the measurement. This would be particularly useful for LiF:Mg,Cu,P in the transition region between the standard TL measurements (< 1 kGy) and the UHTR measurements, which below about 10 kGy show rather high uncertainty (Obryk *et al.*, 2011b). Additionally, such RPL measurement might be exploited for correct setting of the sensitivity of the TL reader. This is an important and uneasy task e.g. in measurements around high-energy accelerators, when doses in one set of detectors may vary from miligrays to hundreds of kilograys (dynamic of thermoluminescence intensity highly exceeds the dynamic range of photomultiplier tubes). Therefore the application of combined RPL/TL measurements seems to be a potential improvement of high-dose dosimetry.

## ACKNOWLEDGEMENTS

The work was supported by the National Centre for Research and Development (Contract No. PBS1/A9/4/2012). Anna Mrozik has been partly supported by the EU Human Capital Operation Program, Polish Project No. POKL.04.0101-00-434/08-00.